\documentclass[twocolumn,prl,superscriptaddress,showpacs]{revtex4}
\usepackage{amsmath,amssymb,graphicx}

\begin{document}

\setlength{\tabcolsep}{1.5mm}
\setcounter{totalnumber}{4}
\setcounter{topnumber}{4}

\setlength{\voffset}{-0cm}
\setlength{\hoffset}{-0.cm}
\addtolength{\textheight}{1.1cm}

%%%%%%%%%%%%%%%%%%%%
%% PRIVATE MACROS %%
%%%%%%%%%%%%%%%%%%%%

\newcommand{\figwidth}{0.90\columnwidth}
\newcommand{\eq}[1]{Eq.(\ref{#1})}
\newcommand{\fig}[1]{Fig.~\ref{#1}}
\newcommand{\sect}[1]{Sec.~\ref{#1}}
\newcommand{\avg}[1]{{\langle #1 \rangle}}
\newcommand{\olcite}[1]{Ref.~\onlinecite{#1}}

%%%%%%%%%%%%%%%%%%%%%%%%%%
%% DOCUMENT STARTS HERE %%
%%%%%%%%%%%%%%%%%%%%%%%%%%

\title{Importance of Metastable States \\ in the Free Energy Landscapes of Polypeptide Chains}

\author{Stefan Auer}
\affiliation{University Chemical Laboratory, Lensfield Road,
Cambridge CB2 1EW, United Kingdom}

\author{Mark A.~Miller}
\affiliation{University Chemical Laboratory, Lensfield Road,
Cambridge CB2 1EW, United Kingdom}

\author{Sergei V.~Krivov}
\affiliation{Laboratoire de Chimie Biophysique, ISIS, Universit{\'e} Louis Pasteur, 67000
Strasbourg, France}

\author{Christopher M.~Dobson}
\affiliation{University Chemical Laboratory, Lensfield Road,
Cambridge CB2 1EW, United Kingdom}

\author{Martin Karplus}
\affiliation{Laboratoire de Chimie Biophysique, ISIS, Universit{\'e} Louis Pasteur, 67000
Strasbourg, France}
\affiliation{Department of Chemistry and Chemical Biology, Harvard University, Cambridge, MA 02138}

\author{Michele Vendruscolo}
\affiliation{University Chemical Laboratory, Lensfield Road,
Cambridge CB2 1EW, United Kingdom}


\date{\today}

\begin{abstract}
We show that the interplay between excluded volume effects,
hydrophobicity, and hydrogen bonding 
of a tube-like representation of a polypeptide chain
gives rise to free energy landscapes that exhibit a small
number of metastable minima corresponding to common structural
motifs observed in proteins.  The complexity of the landscape
increases only moderately with the length of the chain.
Analysis of the temperature dependence
of these landscapes reveals that the stability of specific metastable
states is maximal at a temperature close to the mid-point of folding.
These mestastable states are therefore likely to be of particular significance 
in determining the generic tendency of proteins to aggregate
into potentially pathogenic agents.
\end{abstract}

\maketitle

The mechanism by which proteins fold reliably into their native states
is frequently described by using the concept of a free energy
landscape (FEL) \cite{dill97,dinner00,shea01,onuchic04}.  Under
physiological conditions, the region of configuration space associated
with the native state has the lowest free energy and is therefore
thermodynamically the most stable.  
In addition to the native and fully unfolded states,
intermediate structures have been detected in the folding and 
misfolding processes of many proteins
\cite{radford92,rumbley01,khan04,korzhnev04short}.  
These metastable states can play an important role in the folding process, 
or be kinetic traps that interfere with correct folding.  
The experimental characterization of these states remains challenging,
as they are often transient or disordered, but significant
progress in this direction has recently been
made by combining experiment with theory
\cite{mayor03short,korzhnev04short,gsponer06short}.  
Since specific metastable states can increase the probability of 
misfolding and aggregation \cite{booth97,mcparland02},
it is important to understand the mechanism of their formation.
Indeed, a global characterization of the FEL is crucial to a
full understanding of the relationship between the native 
and metastable states, and the interplay between folding and misfolding.

A complete characterization of the FEL by computational methods 
requires that the free energies of the native and all non-native
structures be calculated.  Two problems arise in this exercise: the
technical issue of sampling a vast number of
configurations separated by a variety of free energy barriers, 
and the conceptual
difficulty of choosing an appropriate set of coordinates to
describe the FEL \cite{dobson98}.  Sampling is typically performed in
the vicinity of an ensemble of folding or unfolding pathways, and the
resulting free energy is then projected on to a subspace defined by one
or two order parameters \cite{shea01}.  For the resulting surfaces to
provide insight into thermodynamics and dynamics, it is crucial that
the chosen order parameters are able to detect the relevant details of
the FEL.  In cases where a long dynamic trajectory that explores a
large region of configuration space is available, it is possible to
mitigate the order parameter problem by using the dynamics to cluster
the configurations and disconnectivity graphs to represent the results
\cite{krivov04}.  To maintain the dynamic approach but improve the
extent of sampling, methods have been developed
that start from a survey of local minima on the {\em potential} 
energy landscape \cite{krivov02}.  
A related approach has been used 
in which potential energy minima are
grouped together by a dynamic criterion based on an approximate rate
theory before their combined free energies are calculated \cite{carr05}.

The level of detail in which a FEL can be computed depends on the
choice of protein model.  For fully atomistic representations,
analysis has been restricted mostly to peptides and small proteins
\cite{dinner00,shea01}.  Coarse-grained descriptions 
allow larger regions of the conformational space to be explored.
In the most tractable models proteins are confined to a
lattice \cite{sali94}. A great deal of valuable insight has
been gained from this approach, but there are limits to how realistic it
can be made.  A promising new model has recently been proposed 
whose distinctive
feature is that the protein backbone is assigned a finite thickness to
account in an effective way for the volume occupied by the amino acid
side chains \cite{hoang04,banavar06short,auer07}.  
The interactions considered include directional hydrogen
bonding (with well depth $e_{\rm HB}$), a local bending stiffness (defined
by an energy penalty $e_{\rm S}$), and pairwise attractive hydrophobic forces
(with energy $e_{\rm W}$).  The protein is thus regarded as a semi-flexible
tube whose radial symmetry is broken by the restraints
imposed by the hydrogen bonds.  The excluded volume of the tube makes
this model significantly different from other off-lattice
coarse-grained models such as beads-on-strings, and also from G\=o
models because it includes no explicit energetic bias towards a
predetermined structure.

The zero-temperature phase diagram of a
24-residue homopolymer was characterised by using this approach
as a function of hydrophobic strength and stiffness, 
each measured relative to the hydrogen bonding strength \cite{hoang04}.  
The resulting set of ground states,
similar to those annotating the figures in this Letter, show a
remarkable resemblance to the structures deposited in the protein data bank
\cite{berman00short}, indicating that the model is capable of
capturing the range of folds available to a real polypeptide chain.  
The directionality of the hydrogen bonding, together with the
excluded volume of the tube (or secondary structure requirements, as
shown recently in a related model \cite{zhang06}) 
confine the energetically
favorable conformations of a polypeptide chain to a small subset of those that
would exist in the absence of such restrictions.

In this Letter, we look beyond the global minima of the potential
energy surfaces, and
characterize the entire FEL of the tube model at finite temperatures.
By repeating the calculations for different values of the 
parameters we examine the effect of hydrophobicity and stiffness on
both native and metastable states.  We also determine how the
relationship between the various structures on the FEL changes with
temperature, which has important implications for understanding 
the competition between folding, misfolding and aggregation.

Projecting a FEL on to any set of order parameters
runs the risk of concealing or distorting features of the folding
process.  To introduce as little prejudgement as possible, we start by
plotting a transition disconnectivity graph (TRDG)
\cite{krivov02,krivov04} in which configurations along a continuous
ergodic trajectory, generated by crankshaft and pivot
Monte Carlo steps, are clustered according to their all-atom
root-mean-square displacement (RMSD) using a threshold of 2.2\AA.  The
trajectory is mapped onto a graph of nodes, each representing a
cluster of configurations, and edges weighted by the number of
transitions between pairs of clusters.  The partition function $Z$ of
a cluster is proportional to its residence time in the simulation,
while that of a transition state between any pair of nodes is given by
the minimum cut that separates them on the graph
\cite{krivov02,krivov04}.  In both cases, the partition function is
related to the free energy by $F/kT=-\ln Z$.  The TRDG is constructed
from these free energies by placing nodes on a vertical free energy
scale and connecting them at a point according to the free
energy of the transition state that connects them.

\begin{figure}
\includegraphics[width=80mm]{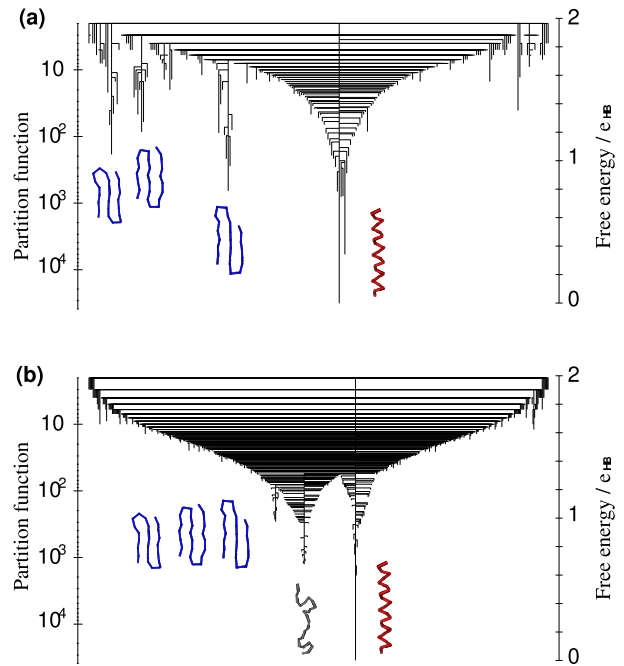}
\caption{\label{trdg} TRDGs for tube parameters $e_{\rm S}=0.4$ and
$e_{\rm W}=0.05$ at temperature $T^*=0.161$.  Configurations are grouped (a) by
RMSD, and (b) by the order parameters $(h,l,n)$.}
\end{figure}

For stiffness $e_{\rm S}=0.4$ and hydrophobicity $e_{\rm W}=0.05$ at
temperature $T^*=0.161$ (where all quantities are specified in units
of $e_{\rm HB}$), a simulation was run for a sufficiently long
time to observe spontaneous folding and unfolding about 100 times.  The
resulting TRDG is shown in Fig.~\ref{trdg}(a).  Most of the branches from
the main stem are vertically short, indicating low barriers and
thus a smooth surface.  The main feature is a well-defined stem
corresponding to a funnel leading to the $\alpha$-helical native state.
Three smaller metastable funnels are also present, corresponding to
three types of three-stranded $\beta$-sheets that differ in
the length of the strands.  Despite the significant time spent in
a variety of unfolded configurations, 
the extended chain does not appear explicitly
in Fig.~\ref{trdg}(a) because the RMSD clustering scheme
splits this state into many sequentially connected groups of
unstructured configurations, in contrast to the structurally
well-defined folded states.

As noted above, the construction of TRDGs relies on 
the availability of an ergodic
dynamical trajectory that samples all the configurations of interest.
However, equilibrium trajectories of computationally feasible lengths
do not sample metastable states correctly on a complex landscape,
where large free energy barriers must be surmounted to reach them.  
To mitigate these problems, biased sampling techniques can be
employed, accepting that the resulting trajectories cannot be
interpreted dynamically.  It is convenient in such simulations to
project the configuration space on to a set of order parameters;  we have chosen
$h$, the number of hydrophobic contacts, and $l$ and $n$, the number
of local and non-local hydrogen bonds, respectively.  The triplet
$(h,l,n)$ can discern the common structural motifs adopted
by the tube.  To test the effect of the projection, we
apply it to the dynamic trajectory depicted in Fig.~\ref{trdg}(a) to
draw a companion TRDG (Fig.~\ref{trdg}(b))
where states are defined by the values of $h$,
$l$ and $n$, rather than by RMSD clustering.  
The most prominent feature in Fig.~\ref{trdg}(b) is again the
funnel of the $\alpha$-helix, but now the three metastable
$\beta$-sheet states appear as one feature, since the order parameters do not
distinguish between them.  The projected graph clearly shows the
metastable free energy minimum of the extended chain, which
corresponds to a large volume of configuration space where $(h,l,n)$
are all close to zero.

Although some differences arise from the choice of the
clustering method, the two TRDGs in \fig{trdg} present a consistent
picture of the FEL.  For example, the total relative partition
function (obtained by summing the contributions of individual
branches) of the three $\beta$-sheet structures in Fig.~\ref {trdg}(a) is
2179, while the corresponding funnel in Fig.~\ref{trdg}(b) sums to a
comparable 2565.  The height of free energy barriers is also somewhat
affected by the projection, but the organization of the graph
according to structural motifs is not.  The
$(h,l,n)$ projection will be used in the biased simulations that follow.

To sample the FEL extensively under conditions where large barriers
exist, we implemented umbrella sampling \cite{torrie74} using a
parabolic potential in $l$ and $n$.  The minimum of the umbrella
potential was placed in turn over points on a grid in these two order
parameters to generate a series of overlapping sampling windows.  Each
run was further enhanced by parallel tempering with four or six stages
of temperature covering a range over which all states from native to
fully unfolded are explored.  The probability histograms $P(h,l,n)$
from all the runs were combined using the multiple histogram
technique \cite{ferrenberg89}, 
enabling the free energy $F(h,l,n)=F_{\rm ref}-kT\ln
P(h,l,n)$ to be determined for a wide range of the order parameters up
to the arbitrary constant $F_{\rm ref}$.

Since $F(h,l,n)$ is a function of three variables, it cannot be depicted by
a contour plot. We therefore adopted a graphical representation that shows
the organization of basins (thermodynamically stable regions) and saddles
(free energy barriers) of $F(h,l,n)$.
A basin is represented by its lowest point, a local minimum of
$F(h,l,n)$ defined by a triplet $(h,l,n)$ for which a unit change in
any individual order parameter or combination of order parameters
leads to a higher free energy.  A saddle point between two basins is
the triplet of lowest free energy from which both basins can be
reached by sequences of downhill stepwise changes in the order
parameters.  In this way, a network of nodes (basins) and edges
(connecting saddles) is built up.  A landscape topology graph (LTG)
was then constructed analogously to a potential energy disconnectivity
graph \cite{becker97}, by placing the nodes on a vertical free energy axis,
and connecting pairs of nodes at the free energy of the highest saddle
on the lowest contiguous path on $F(h,l,n)$ that joins them.
Since each
basin is represented only by its lowest point, the positions of the
nodes and the vertical height of the branches in a LTG cannot be
quantitatively identified with the total free energy of structures and
barriers in the same way as in a TRDG.  However, LTGs reveal particularly
clearly
the major features and organization of the FEL and will be seen in
what follows to be consistent with TRDGs.

\begin{figure}
\includegraphics[width=80mm]{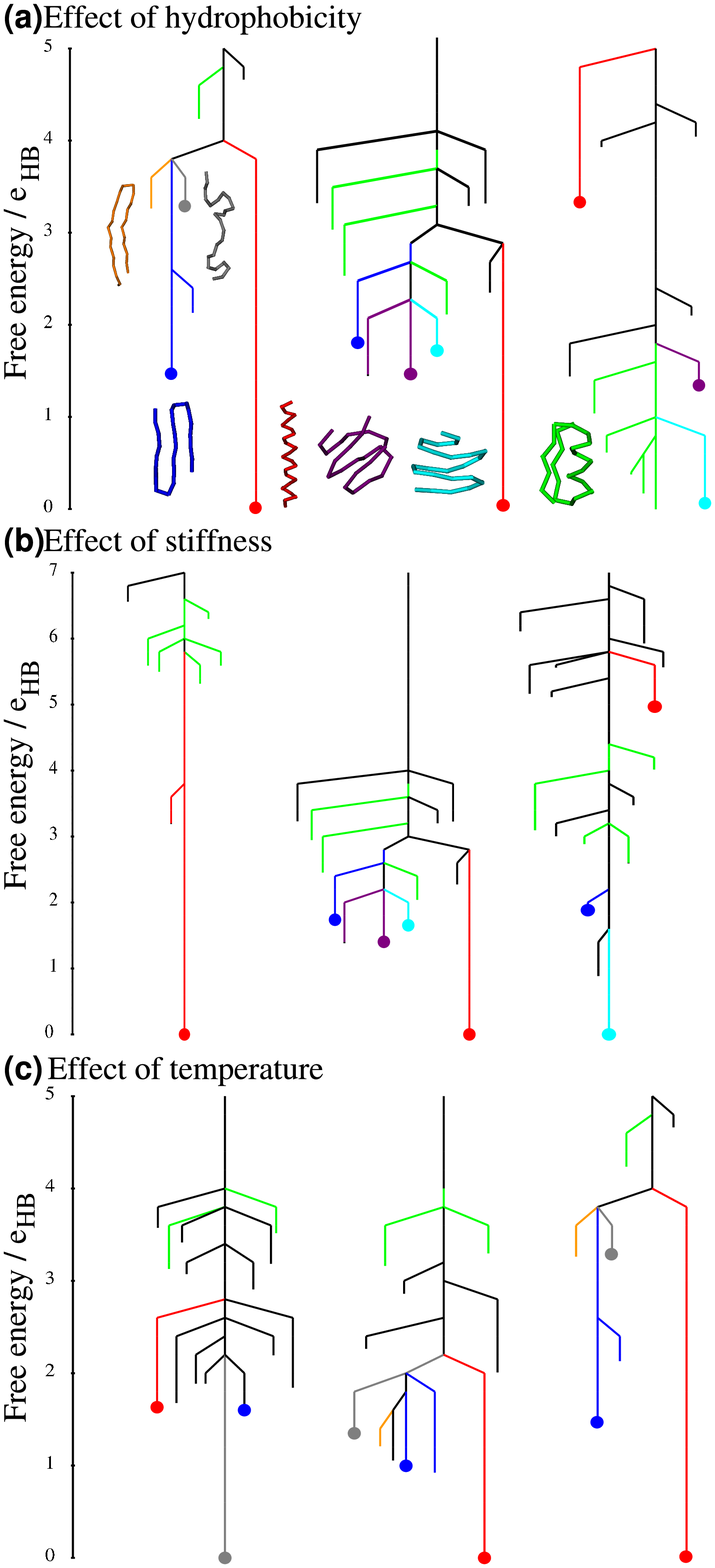}
\caption{\label{trees} LTGs illustrating the effect of: (a) hydrophobicity
(with $e_{\rm S}=0.04$,
$T^*=0.141$ and $e_{\rm W}=0.04,-0.1,-0.2$), (b) chain stiffness (with $e_{\rm
W}=-0.1$, $T^*=0.141$
and $e_{\rm S}=0.2,0.4,0.6$) and (c) temperature (with $e_{\rm S}=0.4$,
$e_{\rm W}=0.05$ and $T^*=0.181,0.16,0.141$).
The branches that correspond to the structural 
motifs depicted in panel (a) are color-coded accordingly.}
\end{figure}

Fig.~\ref{trees} shows LTGs for the 24-residue tube model under a variety
of conditions.  The top-left panel is dominated by a deep free
energy minimum of an $\alpha$-helical state.  The only prominent metastable
structure under these conditions is the $\beta$-sheet, with a
$\beta$-hairpin and the extended chain only weakly metastable.  
The central and right-hand panels of Fig.~\ref{trees} represent
situations in which the hydrophobicity is increased with all other 
parameters held fixed.  The
$\alpha$-helix and $\beta$-sheet are both destabilized with respect to
various $\alpha\beta$ combinations and ultimately the $\beta$-barrel.
By this stage the $\alpha$-helix, though present, lies at such high
free energy that it is unlikely to play a role in equilibrium folding.

Fig.~\ref{trees}(b) shows the effect of increasing stiffness at
constant hydrophobicity.  In all cases, the $\alpha$-helix,
$\beta$-sheet and $\beta$-helix are present.  Altering the stiffness
primarily adjusts the balance of stability between the $\alpha$-helix
and the various $\beta$-strand motifs, while some $\alpha\beta$ combinations
are metastable on each FEL.  
Overall, therefore, the $\alpha$-helix is favored by
low stiffness and hydrophobicity, while competition arises with
$\beta$-strands at higher stiffness, and with more compact structures
at higher hydrophobicity.  
Thus, metastable states are generally present but are small
in number as a result of the energetic and steric restrictions on the
configuration space.

An increase in the number of residues in the polypeptide chain
produces only a modest increment in the complexity
of the FEL.  We found that chains of 
24, 36 and 48 residues 
at $T^*=0.12$, $e_{\rm W}=0.05$ and $e_{\rm S}=0.4$ 
exhibit LTGs (not shown) with 5, 18 and 38 minima, respectively.  
Importantly, the number of
prominent features (long branches) on the graphs remains small,
increasing only from 2 to 4.  
\par
If the tube parameters are held fixed while the temperature is varied,
we see in Fig.~\ref{trees}(c) that
the balance between the free energies of the
$\alpha$-helix and $\beta$-sheet is subtly altered,
but, as expected, the main effect of increasing the temperature is to
stabilize the extended chain with respect to all compact structures.
The LTG in the central panel is drawn at the folding
temperature, where folded and unfolded structures compete equally.

\begin{figure}
\includegraphics[width=80mm]{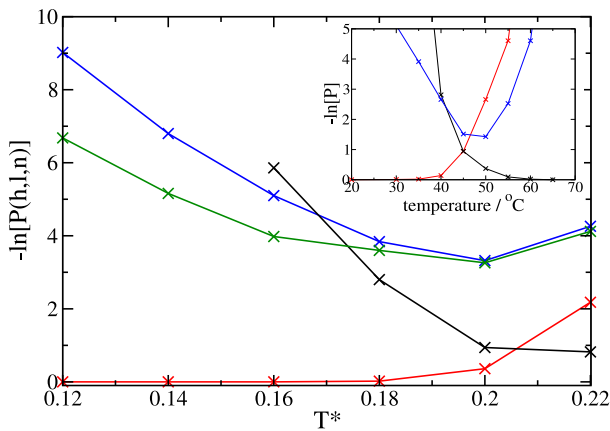}
\caption{\label{stability}
Free energies (structure color code as for Fig.~\ref{trees})
as a function of reduced temperature
for $e_{\rm S}=0.2$ and $e_{\rm W}=0.05$.
Inset: comparison with the native (red line), unfolded (black) and intermediate (blue)
states of human lysozyme, where $P$ is the experimentally determined population
from Ref.~\cite{haezebrouck95short}.}
\end{figure}

The temperature-dependent competition is further examined
in Fig.~\ref{stability} in terms of $-\ln P(h,l,n)$ (i.e., free
energy in units of the associated thermal energy), for the $(h,l,n)$
of lowest free energy in each structural type.  As already seen in
Fig.~\ref{trees}(c), the native $\alpha$-helix is gradually
destabilized with respect to the extended chain with increasing
temperature.  However, the stabilities of the $\beta$-sheet and
$\alpha\beta$ combinations change nonmonotonically.  Starting from low
temperatures, their thermal populations first rise, but, like those of
all compact structures, then fall because of the high entropy of the
extended chain.  Non-native states thus exhibit 
maximum stability close to the folding temperature
while always remaining metastable. 
The experimentally-determined temperature dependence of the stability 
of an amyloidogenic intermediate state of lysozyme 
(inset of Fig.~\ref{stability})
\cite{haezebrouck95short} closely resembles the competition between stable,
metastable and extended states shown in Fig.~\ref{stability}, thus linking
our results to the misfolding process of
proteins, which is often accompanied by pathogenic aggregation
\cite{haezebrouck95short,booth97ashort}.  

The conceptual picture that emerges from the present work is that the
characteristic FEL of a polypetide chain is
dominated by a particular family of related structures, and that
additionally contains a small number of metastable states.
Our results also indicate that there are specific conditions
under which these metastable states are likely to play a major
role in determining the balance between folding and misfolding.


\begin{thebibliography}{27}
\expandafter\ifx\csname natexlab\endcsname\relax\def\natexlab#1{#1}\fi
\expandafter\ifx\csname bibnamefont\endcsname\relax
  \def\bibnamefont#1{#1}\fi
\expandafter\ifx\csname bibfnamefont\endcsname\relax
  \def\bibfnamefont#1{#1}\fi
\expandafter\ifx\csname citenamefont\endcsname\relax
  \def\citenamefont#1{#1}\fi
\expandafter\ifx\csname url\endcsname\relax
  \def\url#1{\texttt{#1}}\fi
\expandafter\ifx\csname urlprefix\endcsname\relax\def\urlprefix{URL }\fi
\providecommand{\bibinfo}[2]{#2}
\providecommand{\eprint}[2][]{\url{#2}}

\bibitem[{\citenamefont{Dill and Chan}(1997)}]{dill97}
\bibinfo{author}{\bibfnamefont{K.~A.} \bibnamefont{Dill}} \bibnamefont{and}
  \bibinfo{author}{\bibfnamefont{H.~S.} \bibnamefont{Chan}},
  \bibinfo{journal}{Nat. Struct. Biol.} \textbf{\bibinfo{volume}{4}},
  \bibinfo{pages}{10} (\bibinfo{year}{1997}).

\bibitem[{\citenamefont{Dinner et~al.}(2000)\citenamefont{Dinner, \v{S}ali,
  Smith, Dobson, and Karplus}}]{dinner00}
\bibinfo{author}{\bibfnamefont{A.~R.} \bibnamefont{Dinner}},
  \bibinfo{author}{\bibfnamefont{A.}~\bibnamefont{\v{S}ali}},
  \bibinfo{author}{\bibfnamefont{L.~J.} \bibnamefont{Smith}},
  \bibinfo{author}{\bibfnamefont{C.~M.} \bibnamefont{Dobson}},
  \bibnamefont{and} \bibinfo{author}{\bibfnamefont{M.}~\bibnamefont{Karplus}},
  \bibinfo{journal}{Trends Biochem. Sci.} \textbf{\bibinfo{volume}{25}},
  \bibinfo{pages}{331} (\bibinfo{year}{2000}).

\bibitem[{\citenamefont{Shea and {Brooks 3rd.}}(2001)}]{shea01}
\bibinfo{author}{\bibfnamefont{J.-E.} \bibnamefont{Shea}} \bibnamefont{and}
  \bibinfo{author}{\bibfnamefont{C.~L.} \bibnamefont{{Brooks 3rd.}}},
  \bibinfo{journal}{Annu. Rev. Phys. Chem.} \textbf{\bibinfo{volume}{52}},
  \bibinfo{pages}{499} (\bibinfo{year}{2001}).

\bibitem[{\citenamefont{Onuchic and Wolynes}(2004)}]{onuchic04}
\bibinfo{author}{\bibfnamefont{J.~N.} \bibnamefont{Onuchic}} \bibnamefont{and}
  \bibinfo{author}{\bibfnamefont{P.~G.} \bibnamefont{Wolynes}},
  \bibinfo{journal}{Curr. Opin. Struct. Biol.} \textbf{\bibinfo{volume}{14}},
  \bibinfo{pages}{70} (\bibinfo{year}{2004}).

\bibitem[{\citenamefont{Radford et~al.}(1992)\citenamefont{Radford, Dobson, and
  Evans}}]{radford92}
\bibinfo{author}{\bibfnamefont{S.~E.} \bibnamefont{Radford}},
  \bibinfo{author}{\bibfnamefont{C.~M.} \bibnamefont{Dobson}},
  \bibnamefont{and} \bibinfo{author}{\bibfnamefont{P.~A.} \bibnamefont{Evans}},
  \bibinfo{journal}{Nature} \textbf{\bibinfo{volume}{358}},
  \bibinfo{pages}{302} (\bibinfo{year}{1992}).

\bibitem[{\citenamefont{Rumbley et~al.}(2001)\citenamefont{Rumbley, Hoang,
  Mayne, and Englander}}]{rumbley01}
\bibinfo{author}{\bibfnamefont{J.}~\bibnamefont{Rumbley}},
  \bibinfo{author}{\bibfnamefont{L.}~\bibnamefont{Hoang}},
  \bibinfo{author}{\bibfnamefont{L.}~\bibnamefont{Mayne}}, \bibnamefont{and}
  \bibinfo{author}{\bibfnamefont{S.~W.} \bibnamefont{Englander}},
  \bibinfo{journal}{Proc. Natl. Acad. Sci. USA} \textbf{\bibinfo{volume}{98}},
  \bibinfo{pages}{105} (\bibinfo{year}{2001}).

\bibitem[{\citenamefont{Khan et~al.}(2004)\citenamefont{Khan, Chuang, Gianni,
  and Fersht}}]{khan04}
\bibinfo{author}{\bibfnamefont{F.}~\bibnamefont{Khan}},
  \bibinfo{author}{\bibfnamefont{J.~I.} \bibnamefont{Chuang}},
  \bibinfo{author}{\bibfnamefont{S.}~\bibnamefont{Gianni}}, \bibnamefont{and}
  \bibinfo{author}{\bibfnamefont{A.~R.} \bibnamefont{Fersht}},
  \bibinfo{journal}{J. Mol. Biol.} \textbf{\bibinfo{volume}{333}},
  \bibinfo{pages}{169} (\bibinfo{year}{2004}).

\bibitem[{\citenamefont{{Korzhnev {\em et al.}}}(2004)}]{korzhnev04short}
\bibinfo{author}{\bibfnamefont{D.~M.} \bibnamefont{{Korzhnev {\em et al.}}}},
  \bibinfo{journal}{Nature} \textbf{\bibinfo{volume}{430}},
  \bibinfo{pages}{586} (\bibinfo{year}{2004}).

\bibitem[{\citenamefont{{Mayor {\em et al.}}}(2003)}]{mayor03short}
\bibinfo{author}{\bibfnamefont{U.}~\bibnamefont{{Mayor {\em et al.}}}},
  \bibinfo{journal}{Nature} \textbf{\bibinfo{volume}{421}},
  \bibinfo{pages}{863} (\bibinfo{year}{2003}).

\bibitem[{\citenamefont{{Gsponer {\em et al.}}}(2006)}]{gsponer06short}
\bibinfo{author}{\bibfnamefont{J.}~\bibnamefont{{Gsponer {\em et al.}}}},
  \bibinfo{journal}{Proc. Natl. Acad. Sci. USA} \textbf{\bibinfo{volume}{103}},
  \bibinfo{pages}{99} (\bibinfo{year}{2006}).

\bibitem[{\citenamefont{Booth}(1997)}]{booth97}
\bibinfo{author}{\bibfnamefont{P.~J.} \bibnamefont{Booth}},
  \bibinfo{journal}{Fold. \& Des.} \textbf{\bibinfo{volume}{2}},
  \bibinfo{pages}{85} (\bibinfo{year}{1997}).

\bibitem[{\citenamefont{McParland et~al.}(2002)\citenamefont{McParland,
  Kalverda, Homans, and Radford}}]{mcparland02}
\bibinfo{author}{\bibfnamefont{V.}~\bibnamefont{McParland}},
  \bibinfo{author}{\bibfnamefont{A.}~\bibnamefont{Kalverda}},
  \bibinfo{author}{\bibfnamefont{S.}~\bibnamefont{Homans}}, \bibnamefont{and}
  \bibinfo{author}{\bibfnamefont{S.}~\bibnamefont{Radford}},
  \bibinfo{journal}{Nat. Struct. Biol.} \textbf{\bibinfo{volume}{9}},
  \bibinfo{pages}{326} (\bibinfo{year}{2002}).

\bibitem[{\citenamefont{Dobson et~al.}(1998)\citenamefont{Dobson, \v{S}ali, and
  Karplus}}]{dobson98}
\bibinfo{author}{\bibfnamefont{C.~M.} \bibnamefont{Dobson}},
  \bibinfo{author}{\bibfnamefont{A.}~\bibnamefont{\v{S}ali}}, \bibnamefont{and}
  \bibinfo{author}{\bibfnamefont{M.}~\bibnamefont{Karplus}},
  \bibinfo{journal}{Angew. Chem. Int. Ed.} \textbf{\bibinfo{volume}{37}},
  \bibinfo{pages}{868} (\bibinfo{year}{1998}).

\bibitem[{\citenamefont{Krivov and Karplus}(2004)}]{krivov04}
\bibinfo{author}{\bibfnamefont{S.~V.} \bibnamefont{Krivov}} \bibnamefont{and}
  \bibinfo{author}{\bibfnamefont{M.}~\bibnamefont{Karplus}},
  \bibinfo{journal}{Proc. Natl. Acad. Sci. USA} \textbf{\bibinfo{volume}{101}},
  \bibinfo{pages}{14766} (\bibinfo{year}{2004}).

\bibitem[{\citenamefont{Krivov and Karplus}(2002)}]{krivov02}
\bibinfo{author}{\bibfnamefont{S.~V.} \bibnamefont{Krivov}} \bibnamefont{and}
  \bibinfo{author}{\bibfnamefont{M.}~\bibnamefont{Karplus}},
  \bibinfo{journal}{J. Chem. Phys.} \textbf{\bibinfo{volume}{117}},
  \bibinfo{pages}{10894} (\bibinfo{year}{2002}).

\bibitem[{\citenamefont{Carr and Wales}(2005)}]{carr05}
\bibinfo{author}{\bibfnamefont{J.~M.} \bibnamefont{Carr}} \bibnamefont{and}
  \bibinfo{author}{\bibfnamefont{D.~J.} \bibnamefont{Wales}},
  \bibinfo{journal}{J. Chem. Phys.} \textbf{\bibinfo{volume}{123}},
  \bibinfo{pages}{234901} (\bibinfo{year}{2005}).

\bibitem[{\citenamefont{\v{S}ali et~al.}(1994)\citenamefont{\v{S}ali,
  Shakhnovich, and Karplus}}]{sali94}
\bibinfo{author}{\bibfnamefont{A.}~\bibnamefont{\v{S}ali}},
  \bibinfo{author}{\bibfnamefont{E.~I.} \bibnamefont{Shakhnovich}},
  \bibnamefont{and} \bibinfo{author}{\bibfnamefont{M.}~\bibnamefont{Karplus}},
  \bibinfo{journal}{Nature} \textbf{\bibinfo{volume}{369}},
  \bibinfo{pages}{248} (\bibinfo{year}{1994}).

\bibitem[{\citenamefont{Hoang et~al.}(2004)\citenamefont{Hoang, Trovato, Seno,
  Banavar, and Maritan}}]{hoang04}
\bibinfo{author}{\bibfnamefont{T.~X.} \bibnamefont{Hoang}},
  \bibinfo{author}{\bibfnamefont{A.}~\bibnamefont{Trovato}},
  \bibinfo{author}{\bibfnamefont{F.}~\bibnamefont{Seno}},
  \bibinfo{author}{\bibfnamefont{J.~R.} \bibnamefont{Banavar}},
  \bibnamefont{and} \bibinfo{author}{\bibfnamefont{A.}~\bibnamefont{Maritan}},
  \bibinfo{journal}{Proc. Natl. Acad. Sci. USA} \textbf{\bibinfo{volume}{101}},
  \bibinfo{pages}{7960} (\bibinfo{year}{2004}).

\bibitem[{\citenamefont{{Banavar {\em et al.}}}(2006)}]{banavar06short}
\bibinfo{author}{\bibfnamefont{J.~R.} \bibnamefont{{Banavar {\em et al.}}}},
  \bibinfo{journal}{Phys. Rev. E} \textbf{\bibinfo{volume}{73}},
  \bibinfo{pages}{031921} (\bibinfo{year}{2006}).

\bibitem[{\citenamefont{Auer et~al.}(2006)\citenamefont{Auer, Dobson
  and Vendruscolo}}]{auer07}
\bibinfo{author}{\bibfnamefont{S.} \bibnamefont{Auer}},
  \bibinfo{author}{\bibfnamefont{C.~M.} \bibnamefont{Dobson}},
  \bibinfo{author}{\bibfnamefont{M.}~\bibnamefont{Vendruscolo}},
  \bibinfo{journal}{HFSP J.} \textbf{\bibinfo{volume}{1}},
  \bibinfo{pages}{137} (\bibinfo{year}{2007}).

\bibitem[{\citenamefont{{Berman {\em et al.}}}(2000)}]{berman00short}
\bibinfo{author}{\bibfnamefont{H.~M.} \bibnamefont{{Berman {\em et al.}}}},
  \bibinfo{journal}{Nucleic Acids Res.} \textbf{\bibinfo{volume}{28}},
  \bibinfo{pages}{235} (\bibinfo{year}{2000}).

\bibitem[{\citenamefont{Zhang et~al.}(2006)\citenamefont{Zhang, Hubner,
  Arakaki, Shakhnovich, and Skolnick}}]{zhang06}
\bibinfo{author}{\bibfnamefont{Y.}~\bibnamefont{Zhang}},
  \bibinfo{author}{\bibfnamefont{I.~A.} \bibnamefont{Hubner}},
  \bibinfo{author}{\bibfnamefont{A.~K.} \bibnamefont{Arakaki}},
  \bibinfo{author}{\bibfnamefont{E.}~\bibnamefont{Shakhnovich}},
  \bibnamefont{and} \bibinfo{author}{\bibfnamefont{J.}~\bibnamefont{Skolnick}},
  \bibinfo{journal}{Proc. Natl. Acad. Sci. USA} \textbf{\bibinfo{volume}{103}},
  \bibinfo{pages}{2605} (\bibinfo{year}{2006}).

\bibitem[{\citenamefont{Torrie and Valleau}(1974)}]{torrie74}
\bibinfo{author}{\bibfnamefont{G.~M.} \bibnamefont{Torrie}} \bibnamefont{and}
  \bibinfo{author}{\bibfnamefont{J.~P.} \bibnamefont{Valleau}},
  \bibinfo{journal}{Chem. Phys. Lett.} \textbf{\bibinfo{volume}{28}},
  \bibinfo{pages}{578} (\bibinfo{year}{1974}).

\bibitem[{\citenamefont{Ferrenberg and Swendsen}(1989)}]{ferrenberg89}
\bibinfo{author}{\bibfnamefont{A.~M.} \bibnamefont{Ferrenberg}}
  \bibnamefont{and} \bibinfo{author}{\bibfnamefont{R.~H.}
  \bibnamefont{Swendsen}}, \bibinfo{journal}{Phys. Rev. Lett.}
  \textbf{\bibinfo{volume}{63}}, \bibinfo{pages}{1195} (\bibinfo{year}{1989}).

\bibitem[{\citenamefont{Becker and Karplus}(1997)}]{becker97}
\bibinfo{author}{\bibfnamefont{O.~M.} \bibnamefont{Becker}} \bibnamefont{and}
  \bibinfo{author}{\bibfnamefont{M.}~\bibnamefont{Karplus}},
  \bibinfo{journal}{J. Chem. Phys.} \textbf{\bibinfo{volume}{106}},
  \bibinfo{pages}{1495} (\bibinfo{year}{1997}).

\bibitem[{\citenamefont{{Haezebrouck {\em et~al.}}}(1995)}]{haezebrouck95short}
\bibinfo{author}{\bibfnamefont{P.}~\bibnamefont{Haezebrouck {\em et al.}}},
  \bibinfo{journal}{J. Mol. Biol.}
  \textbf{\bibinfo{volume}{246}}, \bibinfo{pages}{382} (\bibinfo{year}{1995}).

\bibitem[{\citenamefont{{Booth {\em et al.}}}(1997)}]{booth97ashort}
\bibinfo{author}{\bibfnamefont{D.~R.} \bibnamefont{{Booth {\em et al.}}}},
  \bibinfo{journal}{Nature} \textbf{\bibinfo{volume}{385}},
  \bibinfo{pages}{787} (\bibinfo{year}{1997}).

\end{thebibliography}
\end{document}